\documentclass[10pt,aps,prl,twocolumn,superscriptaddress,floatfix,longbibliography]{revtex4-2}

\usepackage{graphicx}
\usepackage{amsfonts}
\usepackage{amssymb}
\usepackage{amsmath}
\usepackage{txfonts}
\usepackage{lipsum}
\usepackage[dvipsnames]{xcolor}
\usepackage{wasysym}
\usepackage[colorlinks=true, allcolors=blue]{hyperref}
\usepackage{bbold}
\usepackage{etoolbox} 
\usepackage[normalem]{ulem}
\usepackage{mathtools} 
\usepackage{multirow} 
\usepackage{hhline}
\usepackage{mathrsfs} 
\usepackage{soul}

\usepackage{letltxmacro}
\LetLtxMacro{\oldsqrt}{\sqrt}
\renewcommand{\sqrt}[2][\mkern8mu]{\mkern-6mu\mathop{}\oldsqrt[#1]{#2}}

\begin{document}

\title{Interconnected Renormalization of Hubbard Bands and Green’s Function Zeros\\ in Mott Insulators Induced by Strong Magnetic Fluctuations}

\author{Evgeny A. Stepanov}
\affiliation{CPHT, CNRS, {\'E}cole polytechnique, Institut Polytechnique de Paris, 91120 Palaiseau, France}
\affiliation{Coll\`ege de France, Universit\'e PSL, 11 place Marcelin Berthelot, 75005 Paris, France}

\author{Maria Chatzieleftheriou}
\affiliation{CPHT, CNRS, {\'E}cole polytechnique, Institut Polytechnique de Paris, 91120 Palaiseau, France}

\author{Niklas Wagner}
\affiliation{Institut f{\"u}r Theoretische Physik und Astrophysik and W{\"u}rzburg-Dresden Cluster of Excellence ct.qmat, Universit{\"a}t W{\"u}rzburg, 97074 W{\"u}rzburg, Germany}

\author{Giorgio Sangiovanni}
\affiliation{Institut f{\"u}r Theoretische Physik und Astrophysik and W{\"u}rzburg-Dresden Cluster of Excellence ct.qmat, Universit{\"a}t W{\"u}rzburg, 97074 W{\"u}rzburg, Germany}

\begin{abstract}
We analyze the role of spatial electronic correlations and, in particular, of the magnetic fluctuations in Mott insulators. A half-filled Hubbard model is solved at large strength of the repulsion $U$ on a two-dimensional square lattice using an advanced 
diagrammatic non-perturbative approach capable of going beyond Hartree-Fock and single-site dynamical mean-field theories. 
We show that at high temperatures the magnetic fluctuations are weak, and the electronic self-energy of the system is mainly local and is well reproduced by the atomic (Hubbard-I) approximation.
Lowering the temperature toward the 
low-temperature magnetically ordered phase, the non-locality of the self-energy becomes crucial in determining the momentum-dispersion of the Hubbard bands and the Green's function zeros. 
We therefore establish a precise link between Luttinger surface, non-local correlations and spectral properties of the Hubbard bands. 
\end{abstract}

\maketitle

Mott insulators are incompressible states of matter in which collective charge excitations cost an energy proportional to the strength of the electron-electron repulsion~\cite{Mott}. 
A well-established description of these systems is based on the single-orbital Hubbard Hamiltonian consisting of electrons hopping on a lattice and reciprocally interacting with a local repulsion $U$. This model represents the simplest description of the parent compounds of high-temperature superconducting copper-oxides
~\cite{dagotto_correlated_1994, lee_doping_2006}. 
Despite its simplicity, getting a complete physical picture of Mott insulators remains a challenge. This is true even if one restricts to comparatively high temperatures, i.e. above the antiferromagnetic transition and, even more surprisingly, to zero doping.  
The insurgence of local magnetic moments, responsible for the Hubbard satellites in the spectrum of Mott insulators, is indeed a many-body phenomenon that cannot be captured by conventional perturbative approaches~\cite{kondo_resistance_1964,georges_dynamical_1996,coleman_introduction_2015,gunnarsson_breakdown_2017,reitner_attractive_2020} and requires non-trivial analysis~\cite{PhysRevLett.126.056403, PhysRevB.105.155151}.
	
In the last years, Mott insulators have been put under special scrutiny for the classification of their topological properties~\cite{pesin_mott_2010, rachel_topological_2010, zhao_failure_2023, wagner_edge_2023, bollmann_topological_2023}. In the large-$U$ limit, one-electron wave functions are not a valid starting point, jeopardizing the conventional analysis of band-structure-based topology~\cite{slager_impurity-bound_2015, kruthoff_topological_2017, bradlyn_topological_2017}. 
Some suggestions of appropriate four-point correlators as a proxy to the topology of interacting systems have been made~\cite{soldini_interacting_2023, herzog-arbeitman_interacting_2022}, but their handling gets rapidly highly complex. Recently, it has been argued that the simpler level of two-point $T$-products, i.e. the single-particle Green's function $G$, can embody the topological nature of Mott insulators~\cite{wagner2023mott, setty_symmetry_2023, setty_topological_2023}. 
The well-known generalized topological invariant proposed separately by Volovik, Gurarie and Zhang~\cite{unruh_quantum_2007, wang_topological_2010, gurarie_single-particle_2011, manmana_topological_2012, wang_simplified_2012} comprises specular terms and derivatives of $G$ and $G^{-1}$. By itself, this structure indicates how poles and zeros of $G$ can both change the value of the invariant. 
In the strong-coupling limit the poles of $G$ are pushed away from the Fermi level while the Green's function zeros (GFZ), defined as vanishing eigenvalues of $G$ or equivalently as a divergence of the self-energy, form the Luttinger surface at low energy ~\cite{altshuler_luttinger_1998,dzyaloshinskii_consequences_2003,stanescu_theory_2007,rosch_breakdown_2007,dave_absence_2013,fabrizio_emergent_2022,worm_fermi_2023,blason_unified_2023,fabrizio_spin-liquid_2023}.

Hence, the existence and the properties of the Luttinger surface crucially depend on the momentum dependence of the self-energy.
Deep in the Mott phase, the level of knowledge considered to be sufficient has typically stopped at the crude description of dispersionless GFZ, with a few notable exceptions \cite{stanescu_theory_2007,sakai_evolution_2009, sakai_doped_2010,pudleiner_momentum_2016}. These works have not only added temporal fluctuations from single-site dynamical mean field theory (DMFT) \cite{georges_dynamical_1996} but they have also considered the ${\bf k}$-dependence of the self-energy by means of its cluster as well as diagrammatic extensions. 
The full momentum structure of the self-energy is analytically known only in the special case of the exactly solvable local-in-${\bf k}$ Hatsugai-Kohmoto interaction \cite{hatsugai_exactly_1992,mai_14_2023,setty_symmetry_2023,manning-coe_ground_2023}.  
For the Hubbard repulsion, which is instead local in real space, various approximate forms in the Mott insulating case have been previously proposed \cite{PhysRevLett.97.136401, PhysRevB.73.174501,pairault_strong-coupling_2000}.
Yet, the main focus so far has  primarily been on the properties of the Luttinger surface and on the pseudogap formation at finite doping \cite{stanescu_fermi_2006,stanescu_theory_2007,sakai_doped_2010}. We want instead to gain knowledge on the entire energy spectrum, from the Luttinger surface at low frequencies all the way up to the region of the Hubbard bands. 
\begin{figure*}
\includegraphics[width=0.49\linewidth]{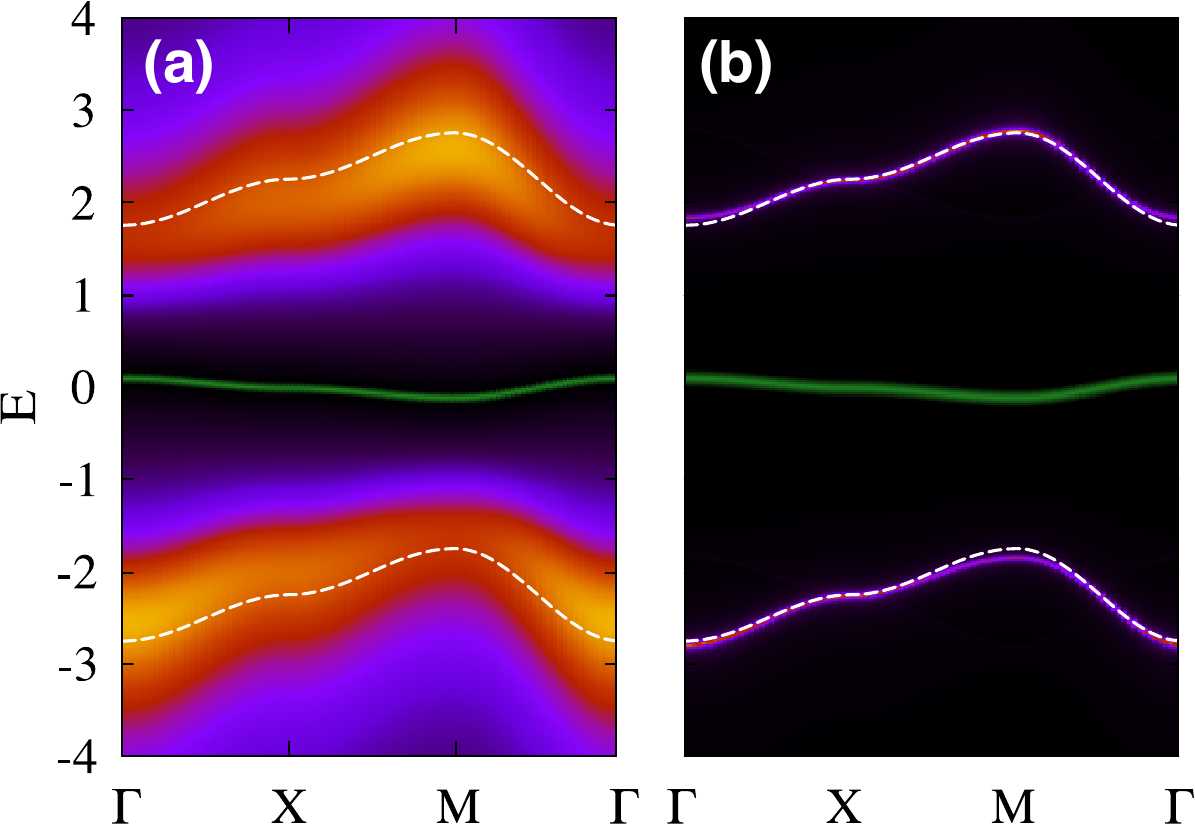} ~~
\includegraphics[width=0.49\linewidth]{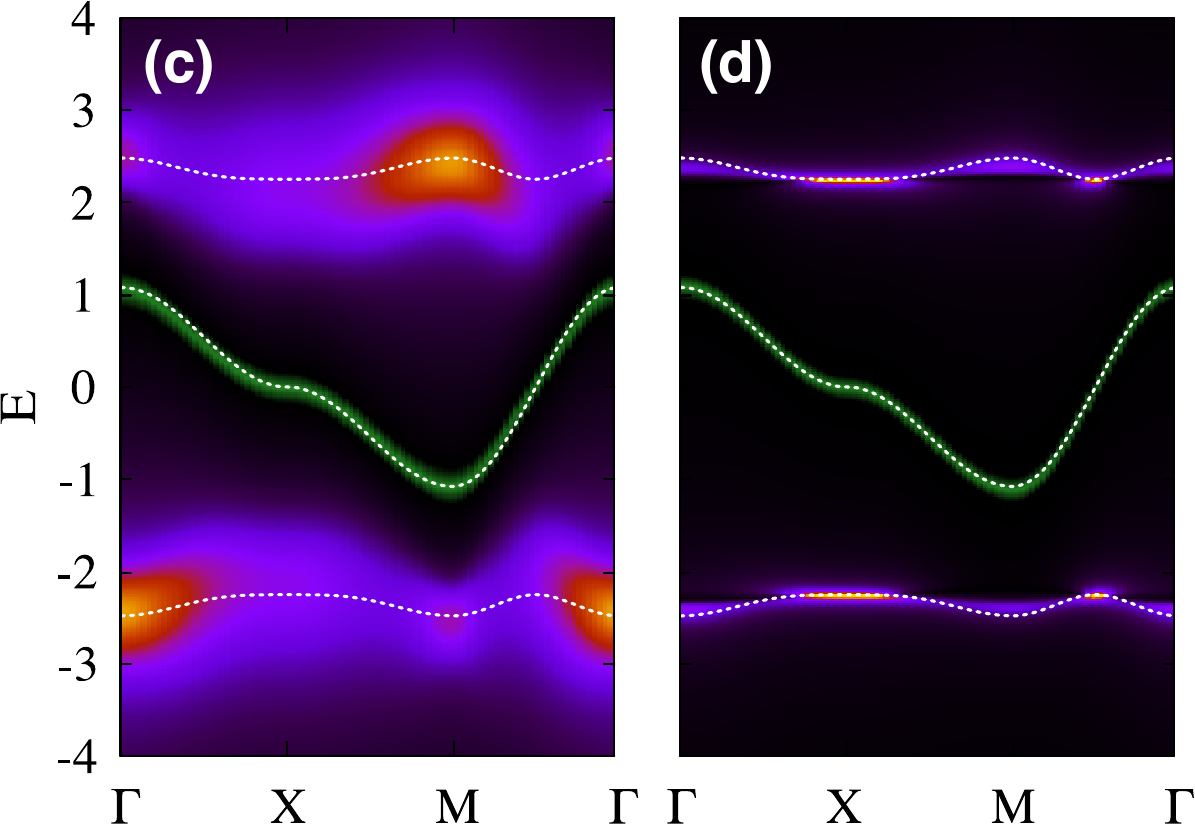} 
\caption{The modulus of the Green's function for a half-filled 2D Hubbard model on a square lattice with the NN hoping $t=0.25$ obtained in the Mott phase (${U=4.5}$) as the function of energy $E$ along the high-symmetry $\Gamma$-X-M-$\Gamma$ path in the BZ.
Panels (a) and (c) correspond to the numerical \mbox{D-TRILEX} calculations performed at ${T=1/5}$ (a) and ${T=1/16}$ (c).
Panels (b) and (d) show the result that corresponds to the analytical expression~\eqref{eq:Sigma_nu} for the self-energy plotted for ${T=1/5}$ (b) and ${T=1/16}$ (d).
Brighter colors correspond to larger intensities.
The GFZ are shown in green color.
Dashed lines in panels (a) and (b) correspond to the atomic limit approximation for the Hubbard bands~\eqref{eq:Spectrum_atom}.
Dotted lines in panels (c) and (d) correspond to the low-frequency approximation for the self-energy~\eqref{eq:Sigma_approx} leading to the analytical expression for the dispersion of the Hubbard bands~\eqref{eq:expression_energy} and of the GFZ~\eqref{eq:expression_zeros}.
\label{fig:GF}}
\end{figure*}

In particular, we pin down the influence of the momentum structure of the GFZ on the dispersion of the Hubbard bands.
This allows us to disentangle a ``trivial'' momentum dependence of the latter, present even with ${\bf k}$-independent GFZ, from the one given by genuine non-local correlations. 
We derive a general expression of the self-energy that not only describes Mott insulators beyond the DMFT level in an accurate way, but also offers a clear interpretation of their physics: spatial magnetic excitations turn out to be essential, not only for the ground state but also for the spectral properties of Mott insulators. 
Using many-particle Feynman diagrammatics based on the dual triply irreducible local expansion (\mbox{D-TRILEX})~\cite{PhysRevB.100.205115, PhysRevB.103.245123, 10.21468/SciPostPhys.13.2.036}, we start from the atomic problem and, upon treating this as a reference system, we account for the leading spatial magnetic fluctuations.
Our main finding is that these fluctuations accounted for beyond the simple DMFT level affect the low- as well as the high-energy scales of the system.
In particular, we establish a one-to-one connection between the renormalization of the GFZ 
and the Hubbard bands by strong magnetic fluctuations.

Even though our results are general, i.e. they are not limited to a specific model, for the sake of definiteness we start from a half-filled 2D Hubbard Hamiltonian on a square lattice:
\begin{align}
\hat{H} = \sum_{jj', \sigma} t_{jj'} c_{j\sigma}^{\dagger} c^{\phantom{\dagger}}_{j'\sigma} + U\sum_{i} n_{j\uparrow} n_{j\downarrow}
\label{eq:Hubbard_model}
\end{align}
written in terms of annihilation (creation) operators $c^{(\dagger)}_{j^{(\prime)}\sigma}$ of electrons on lattice sites $j$ and $j'$ with the spin projection ${\sigma\in\{\uparrow, \downarrow\}}$. 
${n_{j\sigma} = c^{\dagger}_{j\sigma}c^{\phantom{\dagger}}_{j\sigma}}$ is the electronic density operator, $t_{jj'}$ is the hopping amplitude between $j$ and $j'$, and $U$ is the local Coulomb interaction.
Further, we consider a particle-hole symmetric dispersion of electrons determined by the nearest-neighbor (NN) hopping amplitude $t$.
The non-particle-hole-symmetric case is considered as well in the Supplemental Material (SM)~\cite{SM} by additionally introducing the next-NN hopping amplitude $t'$. 
We observe that the derived analytical expression for the self-energy accurately describes the momentum dispersion of the Green's function in both, high- and low-temperatures regimes of the strong-coupling paramagnetic phase. 
Since our approach can be applied also to multi-orbital electronic structures~\cite{10.21468/SciPostPhys.13.2.036, PhysRevLett.127.207205, PhysRevLett.129.096404, PhysRevResearch.5.L022016, stepanov2023charge, stepanov2023orbitalselective}, we propose the use of our analytical expression as a basis for the investigation of topological properties of interacting systems~\cite{wagner2023mott}.

We solve the model~\eqref{eq:Hubbard_model}
in the framework of the
dual perturbation expansion.
The key idea of the method is to replace the non-interacting starting point for the weak-coupling diagrammatic expansion by a suitable interacting reference system. 
After that, a non-perturbative at large $U$ diagrammatic expansion in terms of original degrees of freedom is transformed to a perturbative one in the dual space.
A rigorous derivation of the diagrammatic expansion on the basis of an arbitrary reference system~\cite{BRENER2020168310} has been introduced in the context of the dual fermion (DF)~\cite{PhysRevB.77.033101, PhysRevB.79.045133, PhysRevLett.102.206401, PhysRevB.94.035102, PhysRevB.96.035152}, the dual boson (DB)~\cite{Rubtsov20121320, PhysRevB.90.235135, PhysRevB.93.045107, PhysRevB.94.205110, Stepanov18-2, PhysRevB.100.165128, PhysRevB.102.195109}, and the \mbox{D-TRILEX}~\cite{PhysRevB.100.205115, PhysRevB.103.245123, 10.21468/SciPostPhys.13.2.036, PhysRevLett.127.207205, PhysRevLett.129.096404, PhysRevResearch.5.L022016, stepanov2023charge, stepanov2023orbitalselective, stepanov2021coexisting, Vandelli, Maria} methods.

The advantage of the dual scheme is that it does not result in the double-counting of correlation effects between the reference and remaining problems.
It can be shown, that the total self-energy of the system consists of the two contributions ${\Sigma = \Sigma^{\rm ref} + \overline{\Sigma}}$ (see, e.g., Ref.~\onlinecite{10.21468/SciPostPhys.13.2.036}).
The first term in this expression corresponds to the self-energy of the reference problem $\Sigma^{\rm ref}$.
The second term is the contribution obtained beyond the reference problem using the following exact relation ${\overline{\Sigma} = \tilde{\Sigma} \left[1 + g \tilde{\Sigma}\right]^{-1}}$, where $g$ is the exact Green's function of the reference problem and $\tilde{\Sigma}$ is the self-energy calculated diagrammatically in the dual space (see SM~\cite{SM}).

In this work, all numerical calculations are performed using the \mbox{D-TRILEX} approach with the impurity problem of DMFT~\cite{RevModPhys.68.13} taken as the reference system.
The leading spatial collective electronic  fluctuations are treated diagrammatically beyond DMFT~\cite{RevModPhys.90.025003, Lyakhova_review} in the $GW$-like fashion~\cite{GW1, GW2, GW3} via the renormalized interaction $\tilde{W}^{\rm ch/sp}$ in the charge and spin channels~\cite{PhysRevB.100.205115, PhysRevB.103.245123, 10.21468/SciPostPhys.13.2.036}.
This approach was inspired by the TRILEX method~\cite{PhysRevB.92.115109, PhysRevB.93.235124}. 
\mbox{D-TRILEX} has a much simpler diagrammatic structure but operates at the same level of accuracy as compared to the other dual theories~\cite{PhysRevB.103.245123}. 
As we demonstrate below, this
approach allows us to derive an analytical expression for the self-energy in the strong-coupling regime.

The DMFT impurity problem cannot be solved analytically. 
In order to derive an analytical expression for the self-energy, we require a simpler reference system that, nevertheless, captures the main effects of local correlations.
If the system lies deep in the Mott insulating phase, a sufficient choice for the reference system is an atomic problem, given by the second term in the Hubbard Hamiltonian~\eqref{eq:Hubbard_model}.
In the single-orbital case this reference system can be solved exactly, which gives the following expression for the atomic self-energy ${\Sigma_{\rm at} = U^2/(4i\nu)}$ (see, e.g., Ref.~\cite{ayral:tel-01247625}), where $\nu$ is the fermionic Matsubara frequency. 
The lattice Green's function dressed in the atomic self-energy corresponds to the \mbox{Hubbard-I} approximation~\cite{hubbard1963electron} and reads: 
\begin{align}
G_{{\bf k}\nu} = \frac{i\nu}{(i\nu)^2 - i\nu\varepsilon_{\bf k} - U^2/4},
\label{eq:GF_atom}
\end{align}
where $\varepsilon_{\bf k}$ is the dispersion relation defined by the hopping amplitudes $t_{jj'}$.
The poles of this Green's function define the energy spectrum of the Hubbard bands:
\begin{align}
E = \varepsilon_{\bf k}/2 \pm \sqrt{\varepsilon^2_{\bf k}/4 + U^2/4} = \left(\varepsilon_{\bf k} \pm U\right)/2 + {\cal O}\left(\varepsilon^2_{\bf k}/U\right),
\label{eq:Spectrum_atom}
\end{align}
and the GFZ
${E_{\rm zeros}=0}$ are given by the numerator of Eq.~\eqref{eq:GF_atom}.

In Fig.~\ref{fig:GF}\,(a) we show the numerical result for the modulus of the Green's function obtained for the dispersion ${\varepsilon_{\bf k}=-2t\left(\cos  k_x + \cos k_y \right)}$ defined by the NN hopping ${t=0.25}$ on a square lattice.
The calculations are performed at a rather high temperature ${T=1/5}$ deep in the Mott insulating phase ${U=4.5}$ along the high-symmetry path in the Brillouin zone (BZ) that consists of the ${\Gamma = (0,0)}$, ${\text{X}=(\pi,0)}$, and ${\text{M}=(\pi,\pi)}$ points.
The GFZ are plotted in green color.
They are nearly dispersiveless and can be described by the Hubbard-I approximation ${E_{\rm zeros}=0}$.
The same approximation for the energy spectrum~\eqref{eq:Spectrum_atom} is plotted in white dashed lines.
As expected, it reproduces well the shape of the Hubbard bands in the high-temperature regime of a Mott insulator.

The electronic Green's function changes substantially upon decreasing the temperature. 
Fig.~\ref{fig:GF}\,(c) shows the modulus of the Green's function calculated numerically for the same set of model parameters as in Fig.~\ref{fig:GF}\,(a) but at the low temperature ${T=1/16}$. 
By the low temperature we mean the regime of significant magnetic fluctuations, whose strength enhances upon reducing the temperature.
The strength of the magnetic fluctuations can be assessed by the leading eigenvalue (LE) of the Bethe-Salpeter equation (BSE) for the spin susceptibility~\cite{10.21468/SciPostPhys.13.2.036}.
At ${T=1/5}$ the LE is equal to ${0.22}$, which means that the magnetic fluctuations do not play an important role in this high-temperature regime.
At ${T=1/16}$ the LE is equal to ${0.83}$, which is already close to unity indicating that the magnetic fluctuations are very strong.
Fig.~\ref{fig:GF}\,(c) demonstrates that in the presence of strong magnetic fluctuations the Hubbard bands look different compared to the high-temperature regime of the Mott insulator.
Indeed, the momentum dispersion of the Hubbard bands is no longer described by the Hubbard-I approximation~\eqref{eq:Spectrum_atom} and instead exhibits a mirror-symmetric form with respect to the Fermi energy (${E=0}$) at every ${\bf k}$-point.
In turn, the GFZ
(green color) become very dispersive and their amplitude in energy is several times larger than the amplitude of each Hubbard band. 
It is worth noting that behaviors similar to those reported in Fig.~\ref{fig:GF} have been observed within a completely different approach, namely the Composite Operator Method~\cite{AvellaBook} (see Refs.~\onlinecite{Mancini2004, AvellaEPJB2014} for the two-pole expansion and Refs.~\onlinecite{Avella2007, AvellaAdv2014} for a combination with the non-crossing approximation).

\begin{figure}[t!]
\centering
\includegraphics[width=0.85\linewidth]{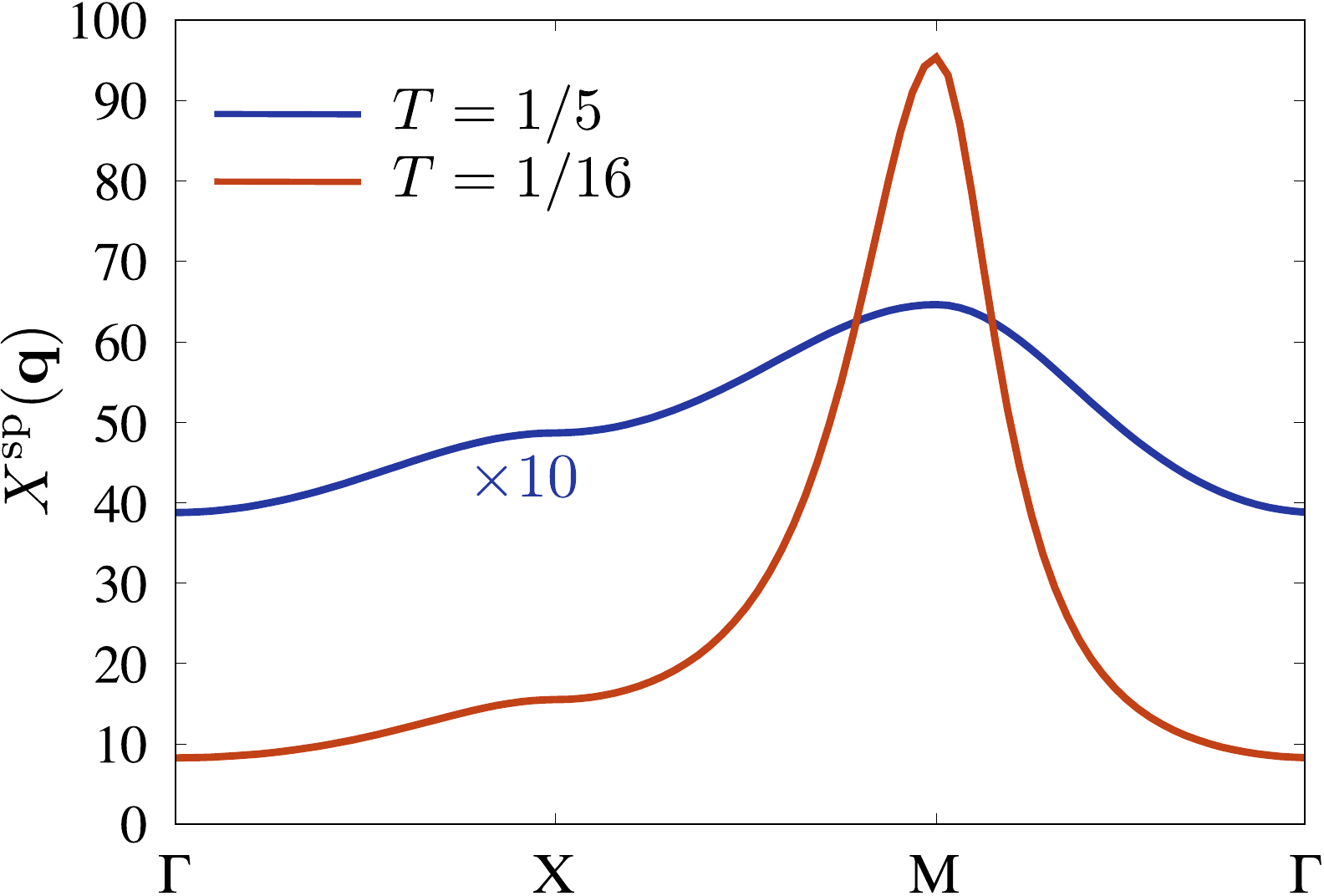}
\caption{The static (${\omega=0}$) spin susceptibility $X^{\rm sp}({\bf q})$ calculated along the high-symmetry path of the BZ in the Mott phase ($U=4.5$) for the two temperatures ${T=1/5}$ (blue) and ${T=1/16}$ (red). The curve of the high $T$ is multiplied by a factor of 10 in order to approach the scale of the low $T$ one. This huge difference of an order of magnitude is due to the effect of strong magnetic fluctuations at low $T$. In this regime, it is valid to assume that the value of ${{\bf Q}=\text{M}}$ dominates the entire susceptibility.
\label{fig:susc_U4.5_b5_b16}}
\end{figure}

In order to account for the effect of magnetic fluctuations we first look at the behavior of the spin susceptibility at different temperatures.
Fig.~\ref{fig:susc_U4.5_b5_b16} shows that in the high temperature regime (${T=1/5}$, blue curve) the spin susceptibility is very small and is not particularly dispersive (in the plot it is even multiplied by a factor of $10$ in order to facilitate a comparison with the low-$T$ data).
Once the temperature is lowered, the spin susceptibility becomes strongly peaked at the ${{\bf Q}=\text{M}}$ point (${T=1/16}$, red curve), which indicates that the leading collective electronic fluctuations in the spin channel are antiferromagnetic (AFM), and the fluctuations at other wave vectors are substantially weaker.
We point out, that, although the spin susceptibility at ${T=1/5}$ also exhibits a maximum at the M point, the difference of this maximum value from the minimum one at the $\Gamma$ point is evidently small.
These observations suggest a simple approximation that can be used in the regime of strong magnetic fluctuations.
Indeed, one can approximate the momentum- and frequency-dependent renormalized interaction $\tilde{W}^{\rm sp}({\bf q},\omega)$ (which is proportional to the spin susceptibility) that enters the \mbox{D-TRILEX} expression for the self-energy, by it's static AFM part, namely ${\tilde{W}^{\rm sp}({\bf q},\omega)\simeq{}\tilde{W}^{\rm sp}({\bf q},\omega)\delta_{{\bf q, Q}}\delta_{\omega, 0}}$.    

Using this approximation in the \mbox{D-TRILEX} diagrammatic expansion together with the atomic problem as a reference system gives us the following expression for the self-energy (see SM~\cite{SM} for a detailed derivation): 
\begin{align}
\Sigma_{{\bf k},\nu} &= \frac{U^2/4 + i\nu\varepsilon_{\bf k+Q}\left(1 - \lambda_\nu \right)}{i\nu - \lambda_\nu\varepsilon_{\bf k+Q}},
\label{eq:Sigma_nu}
\end{align}
where ${\lambda_\nu = \frac{\nu^2+w}{\nu^2+U^2/4}}$ and ${w=3\tilde{W}^{\rm sp}({\bf Q},0)}$.
The energy spectrum of the system is defined by the poles of the Green's function, which is obtained via the usual Dyson equation ${G_{{\bf k},\nu} = \left[i\nu - \varepsilon_{\bf k} - \Sigma_{{\bf k},\nu}\right]^{-1}}$.
The GFZ are given by the denominator of Eq.~\eqref{eq:Sigma_nu}.
The modulus of the Green's function corresponding to the self-energy~\eqref{eq:Sigma_nu} is plotted in Fig.~\ref{fig:GF} for the two temperatures ${T=1/5}$ (b) and ${T=1/16}$ (d).
In these plots the value of $w$ that enters $\lambda_{\nu}$ is deduced from the amplitude of the GFZ 
(green color in Fig.~\ref{fig:GF}\,(a) and (c)) obtained numerically.
We observe that the analytical expression~\eqref{eq:Sigma_nu} for the self-energy perfectly reproduces the momentum dispersion of the Hubbard bands and of the GFZ in both regimes.

When magnetic fluctuations are small (${\lambda_\nu\to0}$) the self-energy~\eqref{eq:Sigma_nu} reduces to the one of the atomic problem, and the relation for the energy spectrum coincides with the \mbox{Hubbard-I} approximation~\eqref{eq:Spectrum_atom} plotted in dashed lines in Fig.~\ref{fig:GF}\,(a) and (b).
Consequently, the GFZ are nearly dispersiveless.
On the contrary, in the low-temperature regime of strong magnetic fluctuations the GFZ are very dispersive.
In order to get an analytical expression for the GFZ,
one can make use of a low-frequency approximation ${\nu^2\ll{}U^2}$ for the self-energy~\eqref{eq:Sigma_nu}, which leads to the following results for the self-energy (${\lambda_\nu \to \lambda=4w/U^2}$):
\begin{align}
\Sigma_{{\bf k},\nu} &\simeq \frac{U^2/4 + i\nu\varepsilon_{\bf k+Q}\left(1 - \lambda\right)}{i\nu - \lambda\varepsilon_{\bf k+Q}}
\label{eq:Sigma_approx}
\end{align}
and the dispersion of the GFZ:
\begin{align}
E_{\rm zeros} = \lambda\varepsilon_{\bf k+Q}.
\label{eq:expression_zeros}
\end{align}
The low-frequency approximation for the self-energy~\eqref{eq:Sigma_approx} also allows one to obtain an analytical expression for the energy:
\begin{align}
E = \frac{\left(\varepsilon_{\bf k} + \varepsilon_{\bf k+Q}\right)}{2} \pm \sqrt{\left(\frac{\varepsilon_{\bf k} + \varepsilon_{\bf k+Q}}{2}\right)^2 + U^2/4 - \lambda\varepsilon_{\bf k}\varepsilon_{\bf k+Q}}.
\label{eq:expression_energy}
\end{align}
Note that, since Eq.~\ref{eq:expression_energy} is justified for ${\nu^2\ll{}U^2}$ only, 
we will apply it cautiously whenever we deal with high-energy Hubbard bands. Yet, this expression turns out to be useful in some illustrative cases also outside its range of validity.

We note that the form of the derived expression for the self-energy~\eqref{eq:Sigma_nu} is different from previous works~\cite{PhysRevLett.97.136401, PhysRevB.73.174501,wagner2023mott} on strong-coupling Hubbard models or quantum spin liquids and disordered spin density wave systems. 
In particular, our expression has an additional contribution to the numerator. 
We find that this contribution is crucial in leading to the mirror-symmetry form of the Hubbard bands, which cannot be achieved without it.
Indeed, in the case of the Hubbard model on a square lattice with the nearest-neighbor hopping amplitude the electronic dispersion satisfies ${\varepsilon_{\bf k+Q} = - \varepsilon_{\bf k}}$, and one gets the following expressions for the energy spectrum~\eqref{eq:expression_energy}: ${E = \pm \sqrt{U^2/4+\lambda\varepsilon^2_{\bf k}}}$. 
The relation for the GFZ~\eqref{eq:expression_zeros}
also simplifies and reads: ${E_{\rm zeros} = - \lambda\varepsilon_{\bf k}}$.
It is important to note, that the renormalization of the energy spectrum and of the GFZ
by magnetic fluctuations are related to each other, because they are determined by the same parameter $\lambda$. 
To illustrate this point, we estimate $\lambda$ from the amplitude of the GFZ 
(green color in Fig.~\ref{fig:GF}\,(c)).
Then, we use this value of $\lambda$ when plotting the dispersion of the Hubbard bands without any further fitting of parameters. 
The analytical result for the energy spectrum~\eqref{eq:expression_energy} and for the dispersion of the GFZ~\eqref{eq:expression_zeros}
is plotted in dotted lines Fig.~\ref{fig:GF}\,(c) and (d).
We observe an excellent agreement between the analytically and numerically obtained results for both the GFZ
and the Hubbard bands.
We find that the strong magnetic fluctuations alter the behaviour of the Hubbard bands, which now exactly follow the dotted lines depicting the analytical expression~\eqref{eq:expression_energy}.
Moreover, the GFZ,
and especially their momentum-dependence, are perfectly reproduced by Eq.~\eqref{eq:expression_zeros}.
Therefore, we find that the dispersion of the Hubbard bands is directly linked to the one of the GFZ.

An advantage of the derived analytical expression for the self-energy~\eqref{eq:Sigma_nu} is that it is not restricted to a particular form of the electronic dispersion and it does not rely on the relation ${\varepsilon_{\bf k+Q} = -\varepsilon_{\bf k}}$, that is not satisfied in a general case.
To illustrate this point, in the SM~\cite{SM} we show the results for a more realistic model with a finite next-nearest-neighbor (NNN) hoping ${t'=-0.3t}$.
Note, that the corresponding dispersion relation ${\varepsilon_{\bf k}=-2t\left(\cos  k_x + \cos k_y \right) + 4t'\cos k_x \cos k_y}$ is not particle-hole symmetric and that ${\varepsilon_{\bf k+Q} \neq -\varepsilon_{\bf k}}$ in this case.
We find that although the dispersion of the GFZ
and of the Hubbard bands differs from the particle-hole-symmetric case considered above, their renormalization is still related to each other through the parameter $\lambda$. 

To conclude, we have studied the momentum-dispersion 
of the energy spectrum and the GFZ in the Mott phase of the Hubbard model. 
We have derived analytical expressions for the self-energy,
which can be applied to 
systems with arbitrary electronic dispersion. 
This fact is confirmed by achieving excellent agreement between analytical and numerical results obtained, both with and without next-nearest-neighbor hopping.
The
generality of the formulas is also manifested in the possibility of calculating a frequency- and momentum-dependent 
Green's function for further considerations for the energy spectrum.
Our rigorously derived equations unveil that a ``parallel'' part of the Hubbard bands comes from the trivial momentum-independent GFZ, while the ``mirror-symmetric'' one arising at lower temperatures is intimately related to the form of the Luttinger surface.
An important remark that emerges from our study is that the amplitudes of the GFZ and the 
Hubbard bands are governed by the same parameter. This is a further confirmation of the consistency of our calculations.
This deeper understanding of the behavior of both the energy spectrum and the GFZ can be employed in studies of topological order in correlated systems, and the analytical formulas can be used in an analogous way to the single-particle quantities typically utilized in the study of non-interacting problems.

\bigskip
\begin{acknowledgments}
The authors would like to thank Alexander Lichtenstein, Silke Biermann, Adolfo Avella and Lorenzo Crippa for useful discussions. 
E.A.S. and M.C. also acknowledge the help of the CPHT computer support team. N.W. and G.S. are supported by
the SFB 1170 Tocotronics, funded by the Deutsche Forschungsgemeinschaft (DFG, German Research Foundation) Project-ID 258499086. G.S. acknowledges financial support from the DFG through the Würzburg-Dresden Cluster of Excellence on Complexity and Topology in Quantum Matter ct.qmat (EXC 2147, project-id 390858490).
\end{acknowledgments}

\bibliography{Ref}
\end{document}


\title{
Supplemental Material\\[0.5cm]
Interconnected Renormalization of Hubbard Bands and Green’s Function Zeros\\ in Mott Insulators Induced by Strong Magnetic Fluctuations
}

\author{Evgeny A. Stepanov}
\affiliation{CPHT, CNRS, {\'E}cole polytechnique, Institut Polytechnique de Paris, 91120 Palaiseau, France}
\affiliation{Coll\`ege de France, Universit\'e PSL, 11 place Marcelin Berthelot, 75005 Paris, France}

\author{Maria Chatzieleftheriou}
\affiliation{CPHT, CNRS, {\'E}cole polytechnique, Institut Polytechnique de Paris, 91120 Palaiseau, France}

\author{Niklas Wagner}
\affiliation{Institut f{\"u}r Theoretische Physik und Astrophysik and W{\"u}rzburg-Dresden Cluster of Excellence ct.qmat, Universit{\"a}t W{\"u}rzburg, 97074 W{\"u}rzburg, Germany}

\author{Giorgio Sangiovanni}
\affiliation{Institut f{\"u}r Theoretische Physik und Astrophysik and W{\"u}rzburg-Dresden Cluster of Excellence ct.qmat, Universit{\"a}t W{\"u}rzburg, 97074 W{\"u}rzburg, Germany}

\maketitle

\section{Analytical expression for the self-energy}

In the framework of dual theories, which are the diagrammatic expansions based on a reference system, the lattice self-energy has the following exact form (see, e.g., Refs.~\cite{RUBTSOV20121320, PhysRevB.93.045107, PhysRevB.94.205110}):
\begin{align}
\Sigma_{{\bf k},\nu} = \Sigma^{\rm ref}_{\nu} + \frac{\tilde{\Sigma}_{{\bf k},\nu}}{1+g_{\nu}\tilde{\Sigma}_{{\bf k},\nu}},
\label{eq:Sigma_latt_SM}
\end{align}
where $\Sigma^{\rm ref}_{\nu}$ and $g_{\nu}$ are the exact self-energy and the Green's function of the reference system, respectively. 
$\tilde{\Sigma}_{{\bf k},\nu}$ is the self-energy in the dual space that accounts for the correlation effect beyond the reference problem. 
${\bf k}$ and $\nu$ are respectively the momentum and fermionic Matsubara frequencies. 
The importance of the denominator in the second term in this expression is discussed in Ref.~\onlinecite{BRENER2020168310}. 

As discussed in the main text, for the case studied in the current work the reference system can be approximated by the atomic problem. The self-energy and the Green's function of this problem are following:
\begin{align}
\Sigma^{\rm ref}_{\nu} = \frac{U^2/4}{i\nu}, ~~~~ g_{\nu} = \frac{-i\nu}{\nu^2 + U^2/4}.
\label{eq:Sigma_g_SM}
\end{align}
Upon substituting these quantities in Eq.~\eqref{eq:Sigma_latt_SM} one gets:
\begin{align}
\Sigma_{{\bf k},\nu} = \frac{U^2/4 + i\nu \, \tilde{\Sigma}_{{\bf k},\nu} \frac{\nu^2}{\nu^2 + U^2/4}}{i\nu + \tilde{\Sigma}_{{\bf k},\nu} \frac{\nu^2}{\nu^2 + U^2/4}}.
\label{eq:Sigma_DT}
\end{align}
The diagrammatic expression for the dual self-energy in the \mbox{D-TRILEX} approximation reads:
\begin{align}
\tilde{\Sigma}_{{\bf k},\nu} = \sum_{{\bf q},\omega,\varsigma} \Lambda^{\hspace{-0.05cm}\varsigma}_{\nu,\omega}\tilde{G}^{\phantom{\varsigma}}_{{\bf k+q},\nu+\omega} \tilde{W}^{\varsigma}_{{\bf q},\omega} \Lambda^{\hspace{-0.05cm}\varsigma}_{\nu+\omega,-\omega}.
\label{eq:Sigma_dual_exact}
\end{align}
Note that this expression has the opposite sign compared to the one introduced in Refs.~\cite{PhysRevB.100.205115, PhysRevB.103.245123, 10.21468/SciPostPhys.13.2.036}. 
The ``$-$'' sign is absorbed in the renormalized interaction so that in the current work ${\tilde{W}^{\varsigma}_{{\bf q},\omega}>0}$.
$\Lambda_{\nu,\omega}$ is the three-point (fermion-boson) vertex function of the reference system in the charge ${(\varsigma={\rm ch})}$ and spin ${(\varsigma={\rm sp}\in\{x,y,z\})}$ channels.
$\tilde{G}_{{\bf k},\nu}$ is the renormalized (dressed) dual fermion Green's function.
For the sake of deriving an analytical expression for the self-energy, we approximate it by the bare (undressed) dual Green's function $\tilde{\cal G}_{{\bf k},\nu}$ that has the following form:
\begin{align}
\tilde{\cal G}_{{\bf k},\nu} = g_{\nu} \left[(\varepsilon_{\bf k} - \Delta_{\nu})^{-1} - g_{\nu} \right]^{-1} g_{\nu}.
\end{align}
Here, $\varepsilon_{\bf k}$ is the dispersion of electrons that in the case of a 2D square lattice with nearest-neighbor hoppings is given by:
\begin{align}
\varepsilon_{\bf k}=-2t\left(\cos  k_x + \cos k_y \right).
\end{align}
We set ${t=0.25}$ so that the half bandwidth ${D=1}$ defines the energy scale of the system.
In the atomic limit the fermionic hybridization function $\Delta_{\nu}$ is identically zero, so one gets:
\begin{align}
\tilde{\cal G}_{{\bf k},\nu} = \frac{g_{\nu} \varepsilon_{\bf k} g_{\nu} }{1 - \varepsilon_{\bf k}g_{\nu}}.
\label{eq:G_dual_SM}
\end{align}

\begin{figure*}[t!]
\centering
\includegraphics[width=0.31\linewidth]{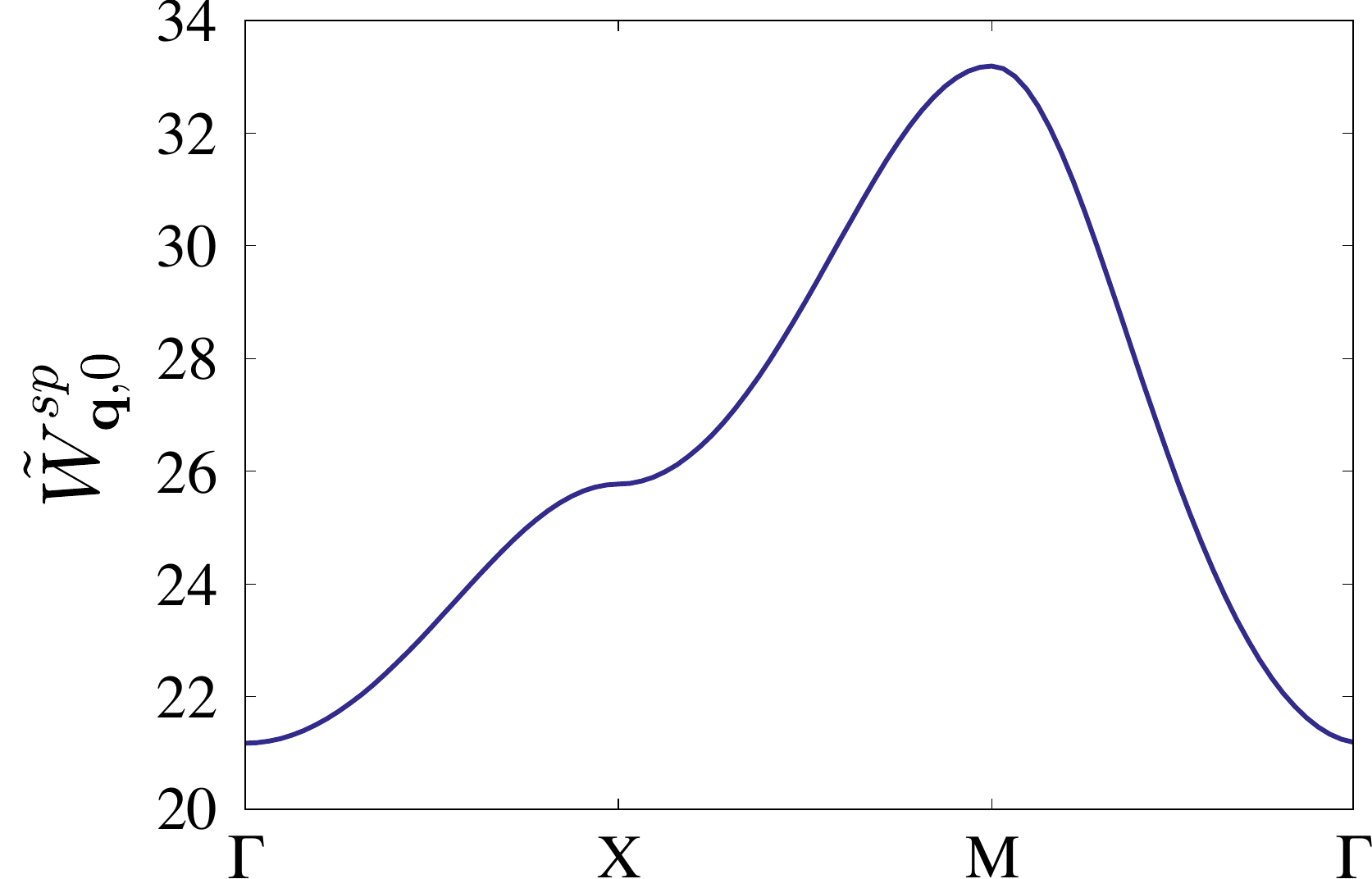} ~~
\includegraphics[width=0.31\linewidth]{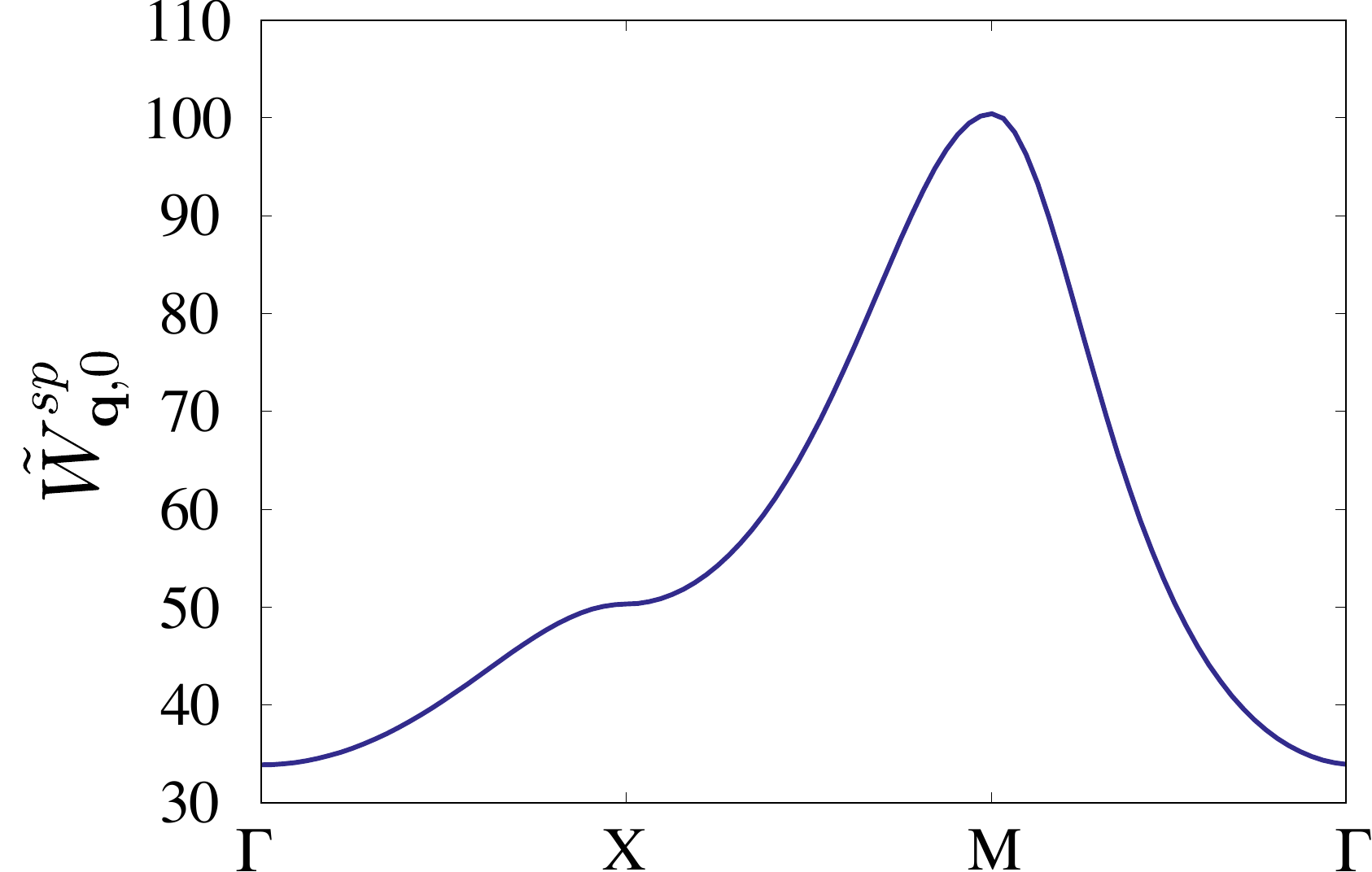} ~~
\includegraphics[width=0.31\linewidth]{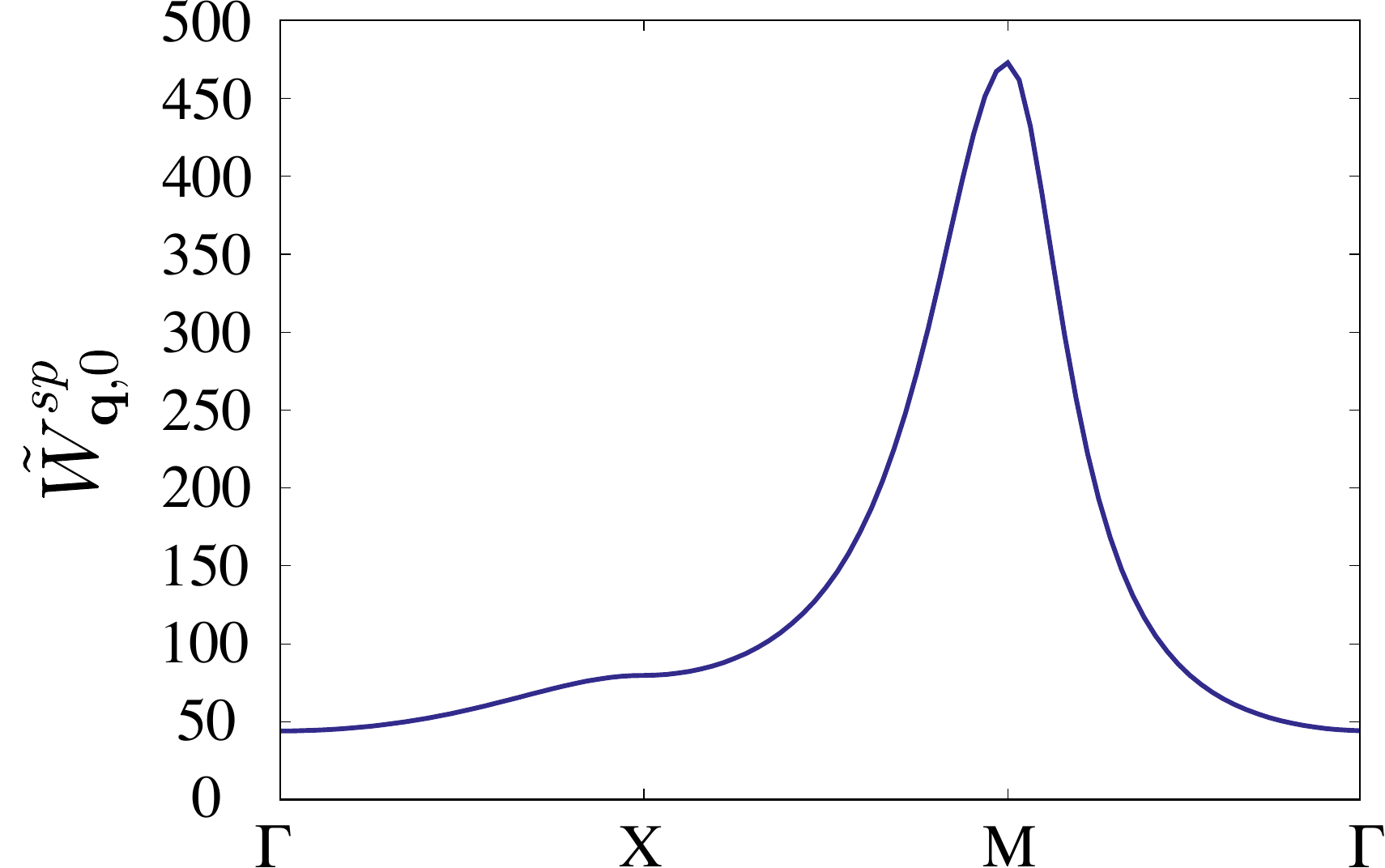}
\caption{The renormalized interaction in the spin channel $\tilde{W}^{\rm sp}_{{\bf q},0}$ obtained using \mbox{D-TRILEX} for ${U=4.5}$ and three different temperatures ${T=1/5}$ (left), ${T=1/10}$ (middle) and ${T=1/16}$ (right). As the temperature is decreased the magnetic fluctuations become stronger the renormalized interaction gets larger and more peaked at the ${{\bf Q}=\text{M}=(\pi,\pi)}$ point. \label{fig:ReSigmaDual10}}
\end{figure*}

One can assume that the main contribution to the dual self-energy comes from the static (${\omega=0}$) part of the renormalized interaction in the spin channel $\tilde{W}^{\rm sp}_{{\bf q},\omega}$ taken at the momentum ${{\bf q=Q}}$, corresponding to the leading magnetic fluctuation (${{\bf Q} = \text{M} =\{\pi,\pi\}}$ in our case). 
In Fig.~\ref{fig:ReSigmaDual10} the interaction $\tilde{W}^{\rm sp}_{{\bf q},0}$ is plotted along the high symmetry path $\Gamma$-X-M-$\Gamma$ in the Mott insulating regime (${U=4.5}$) for three different temperatures. 
Decreasing $T$ (from left to right in Fig.~\ref{fig:ReSigmaDual10}) increases the strength of magnetic fluctuations, leading to the progressive increase of the peak at the M point. The right panel in Fig.~\ref{fig:ReSigmaDual10} thus justifies the above made assumption.
With this approximation the \mbox{D-TRILEX} self-energy simplifies to:
\begin{align}
\tilde{\Sigma}_{{\bf k},\nu} = 3 \Lambda^{\hspace{-0.05cm}\rm sp}_{\nu,0} \tilde{W}^{\rm sp}_{{\bf Q},0} \Lambda^{\hspace{-0.05cm}\rm sp}_{\nu,0}g^2_{\nu} \frac{\varepsilon_{\bf k+Q}}{1 - \varepsilon_{\bf k+Q}g_{\nu}}.
\label{eq:Sigma_dual_Q_SM}
\end{align}
The coefficient ``3'' comes from the sum over the three spin components that are all equal in the paramagnetic case.

The three-point vertex and the susceptibility of the atomic problem are given by the following expressions~\cite{ayral:tel-01247625}: 
\begin{align}
\Lambda^{\rm ch/sp}_{\nu,\omega} &= \frac{1}{1+U^{\rm ch/sp}\chi^{\rm ch/sp}_{\omega}} \Bigg[ 1 + \frac{U^2/4}{i\nu (i\nu + i\omega)} \notag\\
&~~~~+ 
\frac{U}{2} \left(1 + \frac{U^2/4}{\nu^2}\right) \frac{\beta}{2}\left(\tanh\left\{\frac{\beta{}U}{4}\right\}\mp1\right)\delta_{i\omega} \Bigg], \\ 
\chi^{\rm ch/sp}_{\omega} &= -\frac{\beta}{2}\frac{e^{\mp\beta{}U/4}}{\cosh{}\beta{}U/4}\delta_{i\omega}
= \pm\frac{\beta}{2}\left(\tanh\left\{\frac{\beta{}U}{4}\right\}\mp1\right)\delta_{i\omega}.
\end{align}
Note that our definition for the susceptibility differs by a factor of ``-2'' from the definition used in Ref.~\cite{ayral:tel-01247625}.
Thus, the vertex function in the spin channel taken at the zeroth bosonic frequency reads:
\begin{align}
\Lambda^{\rm sp}_{\nu,0} &= \frac{1}{1+U^{\rm sp}\chi^{\rm sp}_{0}} \left[ 1 - \frac{U^2/4}{\nu^2} - \frac{U}{2} \chi^{\rm sp}_{0}\left(1 + \frac{U^2/4}{\nu^2}\right) \right] \notag\\
&= \frac{1}{1+U^{\rm sp}\chi^{\rm sp}_{0}} \left[ 1 - \frac{U^2/4}{\nu^2} + U^{\rm sp} \chi^{\rm sp}_{0}\left(1 + \frac{U^2/4}{\nu^2}\right) \right] \notag\\
&\simeq \frac{\nu^2 + U^2/4}{\nu^2}. 
\end{align}
The dual self-energy becomes:
\begin{align}
\tilde{\Sigma}_{{\bf k},\nu} 
&= -\frac{3 \tilde{W}^{\rm sp}_{{\bf Q},0}}{\nu^2} \frac{\varepsilon_{\bf k+Q}}{1 - g_{\nu}\varepsilon_{\bf k+Q}}. 
\end{align}
Upon substituting this relation for the dual self-energy into Eq.~\eqref{eq:Sigma_DT}
and introducing ${w = 3 \tilde{W}^{\rm sp}_{{\bf Q},0}}$, one gets the expression for the lattice self-energy shown in the main text:
\begin{align}
\Sigma_{{\bf k},\nu} &= \frac{U^2}{4i\nu} - \frac{w\varepsilon_{\bf k+Q}}{\nu^2 + i\nu\varepsilon_{\bf k+Q}\frac{\nu^2+w}{\nu^2+U^2/4}} \notag\\
&= \frac{U^2/4 + i\nu\varepsilon_{\bf k+Q}\left(1 - \frac{\nu^2+w}{\nu^2+U^2/4}\right)}{i\nu - \varepsilon_{\bf k+Q}\frac{\nu^2+w}{\nu^2+U^2/4}} \notag\\
&= \frac{U^2/4 + i\nu\varepsilon_{\bf k+Q}\left(1 - \lambda_\nu\right)}{i\nu - \lambda_\nu\varepsilon_{\bf k+Q}},
\label{eq:Sigma_nu_SM}
\end{align}
where ${\lambda_\nu = \frac{\nu^2+w}{\nu^2+U^2/4}}$.
The low-frequency approximation for the self-energy can be obtained assuming that ${\nu^2\ll{}U^2}$, which gives the following result for the lattice self-energy:
\begin{align}
\Sigma_{{\bf k},\nu} &\simeq \frac{U^2/4 + i\nu\varepsilon_{\bf k+Q}\left(1 - \lambda\right)}{i\nu - \lambda\varepsilon_{\bf k+Q}},
\label{eq:Sigma_approx}
\end{align}
where we introduced ${\lambda=4w/U^2}$.
We note that, up to a constant contribution, the renormalized interaction $\tilde{W}_{{\bf q},\omega}$ can be rewritten through the lattice susceptibility $X^{\varsigma}_{{\bf q},\omega}$ as:
\begin{align}
\tilde{W}^{\varsigma}_{{\bf q},\omega} \simeq U^{\varsigma} X^{\varsigma}_{{\bf q},\omega} U^{\varsigma},
\label{eq:W_X_SM}
\end{align}
where ${U^{\rm ch/sp}=\pm{}U/2}$. 
This results in $\lambda \simeq 3 X^{\rm sp}_{{\bf Q},0}$. 

\begin{figure}[b!]
\centering
\includegraphics[width=0.95\linewidth]{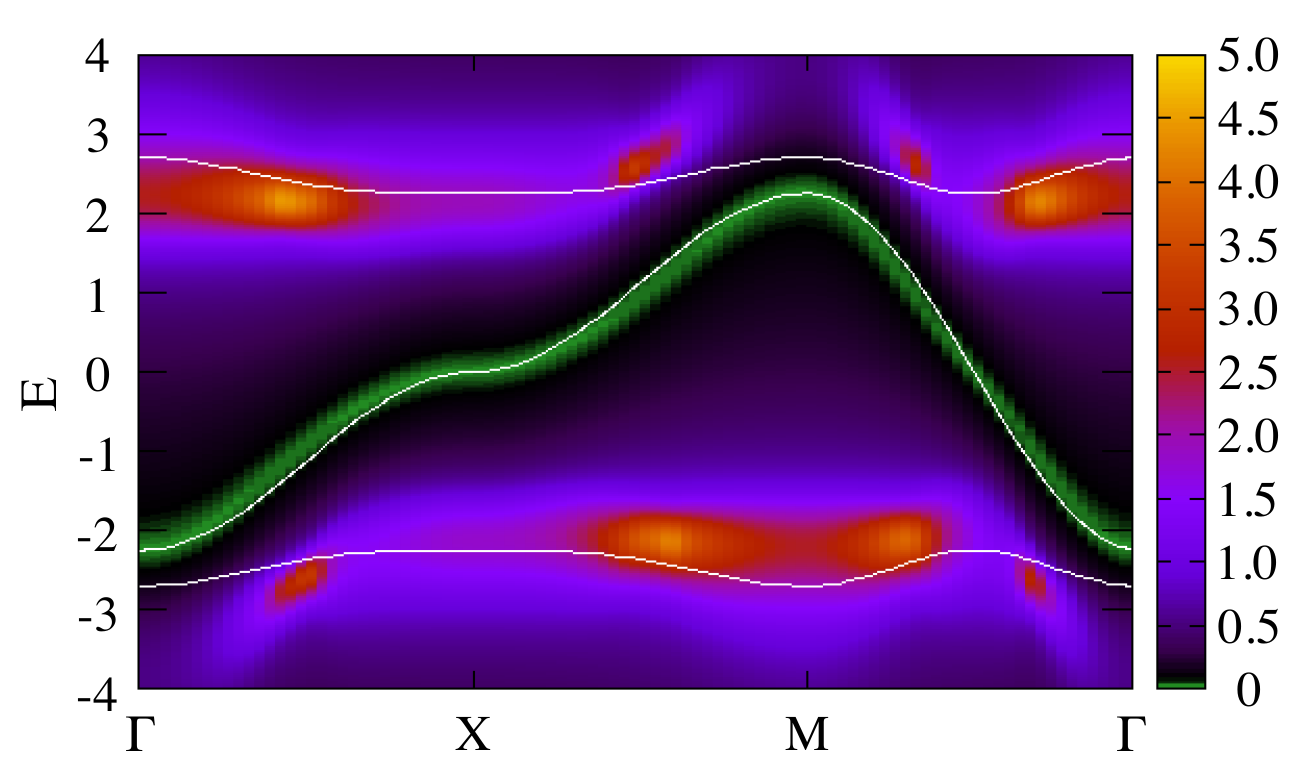} \\
\includegraphics[width=0.95\linewidth] {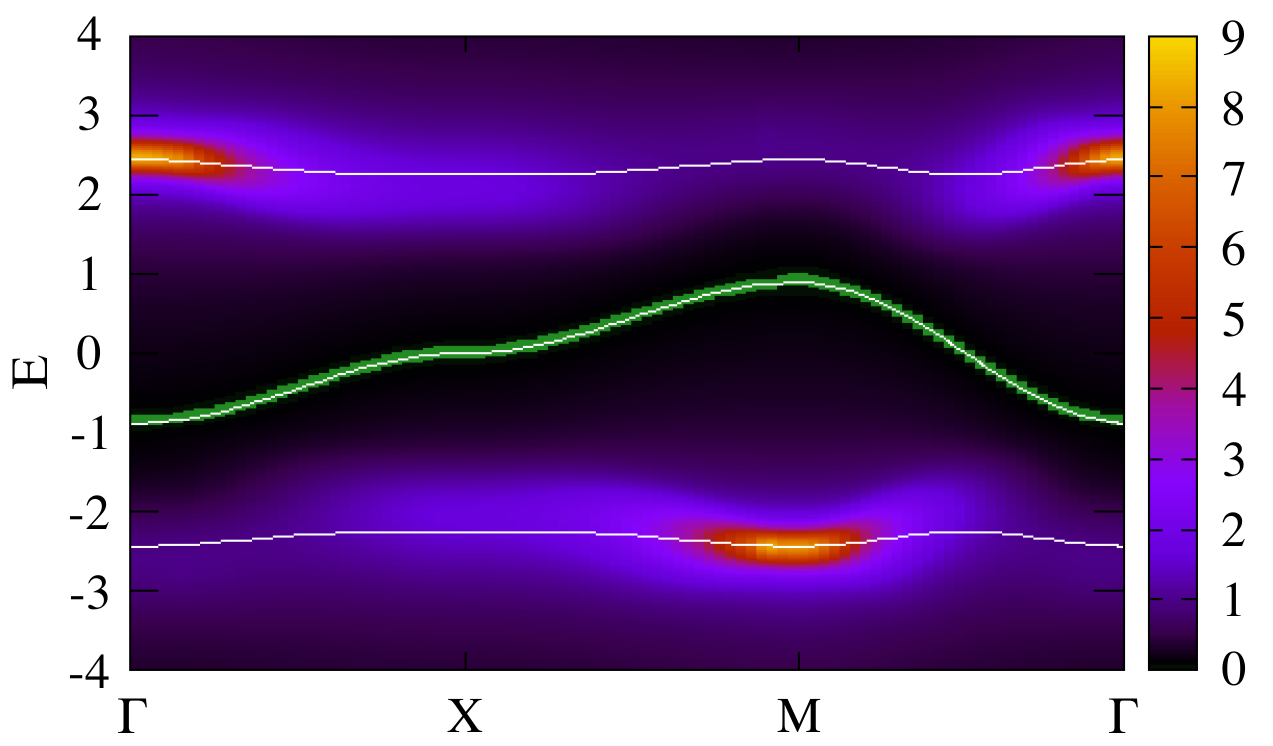} 
\caption{Comparison between single-shot (top panel) and self-consistent (bottom panel) calculations for the modulus of the Green's function obtained at ${T=1/18}$.
\label{fig:zeros_beta18_selfcons}}
\end{figure}

\section{Effect of the renormalized Green's function}

The above derivation for the self-energy has been performed based on the bare dual Green's function. 
For comparison, the numerical results shown in the main text are obtained within single-shot (non-self-consistent) \mbox{D-TRILEX} calculations, where for a decreasing temperature (${T=1/16}$) we found an excellent agreement between the numerical results and our analytical expressions. 
However, by further decreasing the temperature to ${T=1/18}$ the magnetic fluctuations become very large (${\text{LE}=0.95}$) and they lead to GFZ
with large dispersion that push the Hubbard bands to higher energies at the $\Gamma$ and M points. 
This effect can be seen in the top panel of Fig.~\ref{fig:zeros_beta18_selfcons}.
As a result, the derived analytical expression for the energy spectrum (white lines) does not accurately reproduce the Hubbard bands at these ${\bf k}$-points. 
Nevertheless, in this regime one cannot expect single-shot calculations to be accurate, because strong magnetic fluctuations renormalize the Green's function via the self-energy. 
To illustrate this point, we perform self-consistent \mbox{D-TRILEX} calculations for the same temperature ${T=1/18}$, which leads to a decreased ${\text{LE}=0.83}$. 
This value is now similar to the one for the single-shot calculation at ${T=1/16}$ discussed in the main text. 
A self-consistent renormalization of the Green's function decreases the amplitude of the GFZ,
and the renormalized Hubbard bands remind of those found at $T=1/16$ via single-shot \mbox{D-TRILEX} calculations (bottom panel in Fig.~2).
Consequently, the analytical expression for the energy spectrum with the $\lambda$ parameter deduced from the amplitude of the remormalized GFZ 
again accurately reproduces the renormalized Hubbard bands.

\section{Relation to the previously introduced ans\"atze for the self-energy}

It is interesting to explore how our derived expression for the self-energy can be connected to the ans\"atze derived based on the two-pole $t/U$ expansion~\cite{PhysRevLett.97.136401, PhysRevB.73.174501, wagner2023mott}:
\begin{align}
\Sigma_{t/U} = \frac{U^2/4}{i\nu + \tilde{H}_0}. 
\label{eq:ansatz}
\end{align}
In dual theories, a similar expression for the self-energy is given by Eq.~\eqref{eq:Sigma_DT}.
These equations show that the dual self-energy $\tilde\Sigma$ plays a role of the renormalized dispersion of the GFZ,
namely ${\tilde{H}_0 \simeq \tilde{\Sigma}_{{\bf k},\nu} \frac{\nu^2}{\nu^2 + U^2/4}}$. 
Note that Eqs.~\eqref{eq:Sigma_DT} and~\eqref{eq:ansatz} are not identically the same.
In particular, Eq.~\eqref{eq:Sigma_DT} contains an additional contribution in the numerator, which is important for a correct description of the Hubbard bands (see main text).

One can also connect $\tilde{H}_0$ obtained in the \mbox{D-TRILEX} approach to the static magnetic susceptibility. 
To this aim we do not restrict ourselves to the leading momentum ${{\bf q=Q}}$ in the static renormalized interaction $\tilde{W}^{\rm sp}_{{\bf q},0}$, but instead approximate the dual Green's function by a numerator in Eq.~\eqref{eq:G_dual_SM}:
\begin{align}
\tilde{G}_{\bf{k},\nu} \simeq g^2_{\nu}\varepsilon_{\bf k} = \frac{-\varepsilon_{\bf k}\nu^2}{(\nu^2+U^2/4)^2}.
\end{align}
Then, the dual self-energy~\eqref{eq:Sigma_dual_exact} reduces to:
\begin{align}
\tilde{\Sigma}_{{\bf k},\nu} &= -\frac{3}{\nu^2}\sum_{{\bf q}} \varepsilon_{\bf k+q}\tilde{W}^{\rm sp}_{{\bf q},0}. 
\end{align}
Upon using Eq.~\eqref{eq:W_X_SM}, the renormalized dispersion of the GFZ
becomes:
\begin{align}
\tilde{H}_0 = \tilde{\Sigma}_{{\bf k},\nu} \frac{\nu^2}{\nu^2 + U^2/4} &= -3\frac{U^2/4}{\nu^2 + U^2/4}\sum_{{\bf q}} \varepsilon_{\bf k+q}X^{\rm sp}_{{\bf q},0} \notag\\
&\simeq -3\sum_{{\bf q}} \varepsilon_{\bf k+q}X^{\rm sp}_{{\bf q},0},
\label{eq:Zeros_X}
\end{align}
where we used a low-frequency approximation to get the last line.
This expression reminds of the relation for $\tilde{H}_0$ obtained within the two-pole $t/U$ expansion shown in the Supplemental Material of Ref.~\cite{wagner2023mott}. 

\section{Effect of the next-nearest-neighbor hopping}

\begin{figure}[b!]
\centering
\includegraphics[width=0.95\linewidth]{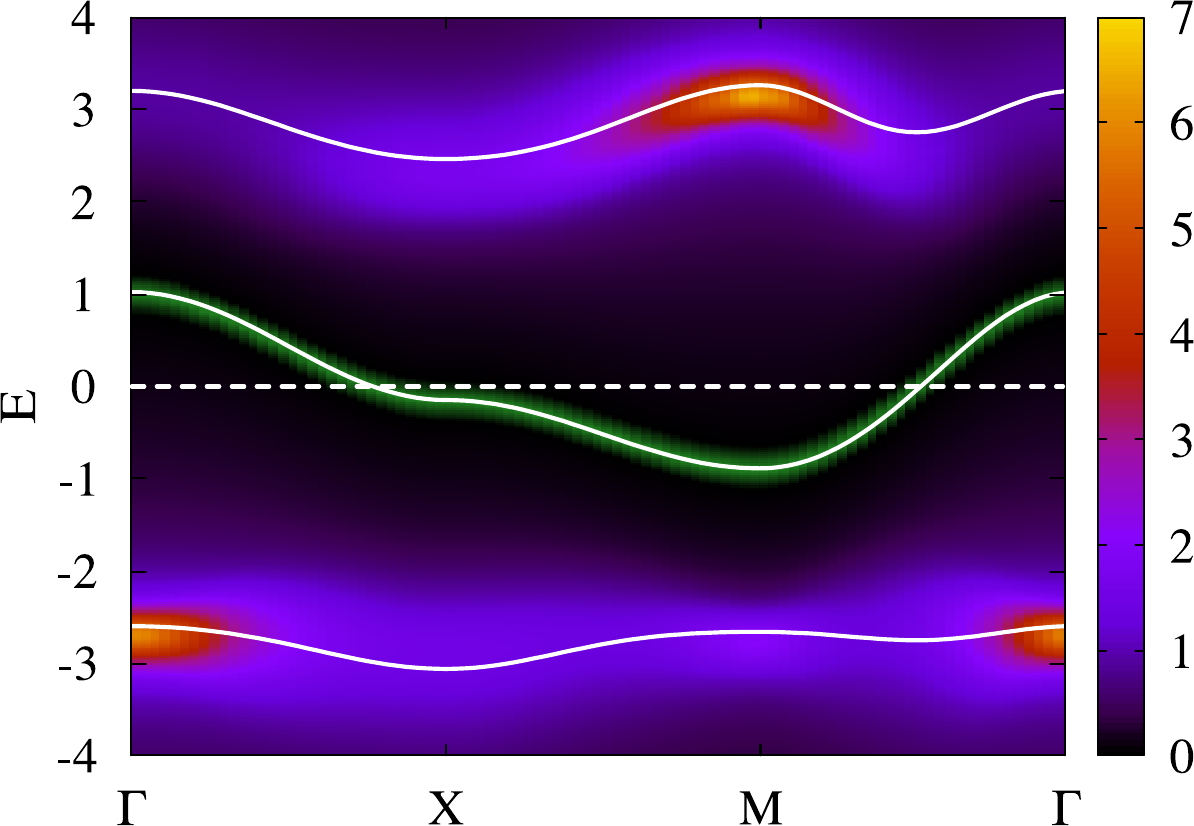}
\caption{The modulus of the Green’s function for a half-filled 2D Hubbard model on a square lattice with the NNN hoping ${t'=0.3t}$. The result is calculated along the high-symmetry path using \mbox{D-TRILEX} for ${U=5.5}$ and ${T=1/22}$. The GFZ
are shown in green. The analytical expressions are plotted in solid white lines. The horizontal dashed line at ${E=0}$ indicates the Fermi energy.
\label{fig:greens_analytic_expression_onplot_NNN}}
\end{figure}

The derived above analytical expression for the self-energy is not restricted to a particular form of the electronic dispersion.
To illustrate this point, let us consider the electronic dispersion ${\varepsilon_{\bf k}=-2t\left(\cos  k_x + \cos k_y \right) + 4t'\cos k_x \cos k_y}$ corresponding to the nearest-neighbor ${t=0.25}$ and the next-nearest-neighbor (NNN) ${t'=-0.3t}$ hoppings on a square lattice.
The case of ${t'=0}$ is considered in the main text.

The presence of the NNN hoping term suppresses the strength of magnetic fluctuations~\cite{PhysRevB.35.3359, PhysRevB.61.2521, JPSJ.71.2109, PhysRevB.81.195122} and also leads to an increased critical value of the Coulomb interaction $U_c$ for the Mott transition~\cite{PhysRevB.74.014421, PhysRevB.77.064427, PhysRevB.88.075114}.
In order to rigorously compare with the results displayed in the main text one needs to perform calculations for a slightly larger value of $U$ and lower temperature.
We find that at ${U=5.5}$ and ${T=1/22}$ the system lies in a region of considerable magnetic fluctuations defined by the same leading eigenvalue ${\text{LE}=0.83}$ as in the ${t'=0}$ case. 
Fig.~\ref{fig:greens_analytic_expression_onplot_NNN} illustrates the modulus of the Green's function obtained using \mbox{D-TRILEX} for this choice of model parameters.
As in the ${t'=0}$ case, we estimate the parameter $\lambda$ from the amplitude of the GFZ
and use this value in the analytical expression for the energy spectrum (Eq.~(7) in the main text).
The result is plotted in solid white lines.
We again find an excellent agreement between the derived analytical expression for the energy spectrum and the numerical data, although in this case the lower and upper Hubbard bands are no longer mirror-symmetric with respect to the Fermi energy at every ${\bf k}$-point and have different forms due to $\varepsilon_{\bf k+Q} \neq -\varepsilon_{\bf k}$.

Estimating $\lambda$ from the amplitude of the GFZ
uncovers yet another interesting effect. 
We find that the dispersion of the GFZ
is actually not fully described by Eq.~(6) in the main text. 
This is clear from the fact that the X point of the GFZ
dispersion is attracted to the Fermi energy, which corresponds to a bit different value of the NNN hopping ${t'= -0.11t}$.
We attribute this feature to the effect of electronic correlations that usually tend to attract and even pin the van Hove singularity of the quasi-particle band to the Fermi energy (see, e.g., Refs.~\cite{PhysRevB.54.12505, PhysRevLett.89.076401, PhysRevB.72.205121, PhysRevB.84.245107, PhysRevB.82.155126}).
Remarkably, although in the Mott insulating case the quasi-particle band is not present at the Fermi energy, it is replaced by the GFZ
for which, apparently, a similar renormalization by the electronic correlations holds.

The attraction of the van Hove singularity to the Fermi energy can be captured, e.g., by the fluctuation-exchange (FLEX) theory~\cite{PhysRevB.54.12505}.
We note that the \mbox{D-TRILEX} self-energy has the FLEX-like diagrammatic form that additionally accounts for the vertex corrections~\cite{PhysRevB.100.205115, PhysRevB.103.245123, 10.21468/SciPostPhys.13.2.036}.
Furthermore, the \mbox{D-TRILEX} self-energy $\tilde{\Sigma}_{{\bf k},\nu}$ gives the dispersion of the GFZ,
as discussed in the previous section below Eq.~\eqref{eq:Sigma_DT}.
These facts allow us to argue that the observed attraction of the van Hove singularity of the GFZ
to the Fermi energy is induced by the same electronic correlations as in the case of the electronic dispersion.
Since Eq.~(6) in the main text does not reproduce the shift of the van Hove singularity seen in the numerical result, this effect goes beyond the static approximation for the renormalized interaction $\tilde{W}$.
Therefore, for a more accurate description of the GFZ
one could use Eq.~\eqref{eq:Sigma_DT} that, however, is much more complex than Eq.~\eqref{eq:Sigma_nu_SM} due to the presence of the dual self-energy $\tilde{\Sigma}_{{\bf k},\nu}$.  

\bibliography{Ref_sup}